\documentclass[prd, amsmath,twocolumn,showpacs,aps,prl]{revtex4-1}
\usepackage{bm}
\usepackage{graphics}
\usepackage{graphicx}
\usepackage{color}
\usepackage[colorlinks=true,linkcolor=blue,pagecolor=blue,filecolor=blue,menucolor=blue,urlcolor=blue,citecolor=blue,anchorcolor=blue]{hyperref}%
\DeclareUnicodeCharacter{2212}{-}
\begin{document}

\title{Superconductivity in dilute system of sites with strong electron-electron attraction}
\author{A.~Yu.~Zyuzin}

\affiliation{A.~F.~Ioffe Physical-Technical Institute of
Russian Academy of Sciences, 194021 St. Petersburg, Russia}


\begin{abstract}
We consider the effect of strong electron-electron attraction on superconductivity in the dilute system of the negative U-centers. They couple to the conducting electronic states and mediate attraction between electrons. We predict the formation of the local Cooper pairs provided attraction exceeds a certain threshold value. 
The global coherent superconducting state appears only due to the Andreev scattering between the centers at low temperatures. 
The dependence of the superconducting transition temperature and the second critical magnetic field on attraction strength is calculated.
\end{abstract}
\maketitle

\section{Introduction}
Electron gas undergoes transition to the superconducting state even at weak attraction between the particles. The formation of electron pairs and their successive condensation takes place at the same temperature. As attraction strength increases, the continuous transition to the state with strongly coupled composite bosons arises \cite{bib:Leggett,bib:Nozieres, bib:Melo}. Narrow electron band structure favors
the two-electron bound states \cite{bib:Micnas}, which are characterized by the pair dissociation temperature. On the low-temperature side, a coherent state similar to the Bose condensate emerges.
Here the coherent transition temperature is determined by the inverse time of inter-particle exchange. For Bose particles, this time is of the order the inter-particle distance flight time  (here and below $\hbar=k_B=1$).

Semiconductors at a low carrier concentration have relatively large values of the superconducting transition temperature. 
An original feature of such systems is the negative U-centers at resonance levels that arise in the main electronic band upon doping with certain impurities. 
The $A^{\mathrm{IV}} B^{\mathrm{VI}}$ compounds \cite{bib:Chernik, bib:Kaidanov,bib:Nemov} doped with the so-called valence skippers \cite{bib:Varma} are particular examples of such semiconductors. 

Superconducting models with local on-impurity attraction were considered in several articles \cite{bib:Mal'shukov}.
The recent theoretical and experimental studies of these compounds can be found in Refs. \cite{bib:Koba, bib:Haldol,bib:Girado-Gallo,  bib:kriener}.
It was noted that the highest superconducting transition temperature can be reached in the regime when the Fermi level is pinned at the resonances provided the degeneracy of energy levels with the number of electrons differs by two $E(N)\simeq E(N+2)$. 
Here $E(N)$ is the energy of the center with $N$ electrons. The presence of such degeneracy raises the question of the influence of the charge Kondo effect on superconductivity \cite{bib:Dzero, bib:Matsushita}.

Using the Hubbard-Stratonovich transformation we study negative U-centers mediated superconductivity at the vicinity as well as far beyond the level degeneracy limit. A weak coupling regime exists up to the emergence of degeneracy and can be described by the BCS approach. As attraction becomes stronger, one notices the resemblance with the problem of composite bosons formation, where Bose pairs are formed locally. Although, the inter-particle exchange of bosons, which leads to the emergence
of a coherent state, occurs in our case due to Andreev scattering. Coherence is established at temperatures much lower than the pair breakdown temperature. We also calculate the upper critical magnetic field. It is shown that in the strong coupling regime magnetic field does not lead to dissociation of bosons, but rather destroys their coherence.

 \section{Main Definitions}
We begin with the system of negative U-centers.
They are formed on the dopants whose electronic resonance states are weakly overlaped with the conducting states. 
The Hamiltonian of the system in terms of electron creation and annihilation operators has a well-known form
\begin{eqnarray}\label{total ham}\nonumber
H &=&\sum_{\textbf{p},\sigma}[ \epsilon(\textbf{p})-\mu]a^{+}(\textbf{p},\sigma)a(\textbf{p},\sigma)+U_{imp}
\\\nonumber
&+&\sum_{i,\sigma,\textbf{p}} V\big[a^{+}(\textbf{p},\sigma)a_i(\sigma)+a_i^{+}(\sigma)a(\textbf{p},\sigma)\big]\\
&+&\sum_{i}\big[(\epsilon_{R}-\mu)(n_{i,\downarrow}+n_{i,\uparrow})
-Wn_{i,\downarrow}n_{i,\uparrow}\big].
\end{eqnarray}
Here 
$a^{+}(\textbf{p},\sigma), a(\textbf{p},\sigma)$ are the creation and annihilation operators of the conducting electrons with spectrum $\epsilon(\textbf{p})$, $\mu$ is the chemical potential, and $U_{imp}$ describes random scattering of the conducting electrons. 
Third term in Hamiltonian (\ref{total ham}) describes hybridization between the conducting and the resonance states. Here $a^{+}_{i}(\sigma), a_{i}(\sigma)$ are  creation and annihilation operators of electrons on the resonance site denoted by index $i$.
The last term in (\ref{total ham}) contains a sum of Hamiltonians, which describe the resonance sites, where $E_{R}\equiv\epsilon_{R}-\mu$ is the resonance energy measured with respect to the chemical potential, and 
$W>0$ is the energy of the electron-electron attraction on site $i$. 


We consider a situation where the resonance site contains one resonance state in the relevant energy region near the Fermi level. Hence the Green function of the resonance site, which is coupled to the band states, might be approximated as 
\begin{equation}\label{g-func}
g_{i}(\textbf{r},\textbf{r}',\omega_{n})=\psi (\textbf{r}-\textbf{r}_{i}) \psi (\textbf{r}'-\textbf{r}_{i})g(\omega_{n}),
\end{equation}
where 
\begin{equation}\label{g1-func}
 g(\omega_{n})=\frac{1}{i\omega_{n}-E_{R}+i\gamma sign(\omega_{n})},
\end{equation}
with the Matsubara frequency $\omega_{n}=(2n+1)\pi T$, the wavefunction of localized state $\psi (\textbf{r})$.
We assume that the wave-function is constant $\sim a^{-3/2}$ in a cube of size $a$ of the order wave-function localization length.
 Smearing of the resonance due to hybridization with the band states is described by
 \begin{equation}
 \gamma=\pi a^{3}\nu_{0}V^{2}, 
\end{equation}
where $\nu_{0}$ is the one-spin density of the conducting states at the Fermi level.

We consider the disordered conducting electrons, where mean free time $\tau$ is determined mostly by scattering on non-resonance sites. The Green function of conducting electrons averaged over the impurities scattering \cite{bib:abricos} is given by
\begin{equation}\label{cond-green}
G(\textbf{r},\textbf{r}',\omega_{n})=\int \frac{d^{3}\textbf{p}}{(2\pi)^{3}}\frac{\exp[i\textbf{p}(\textbf{r}-\textbf{r}')]}{i\omega_{n}-\epsilon(\textbf{p})+\mu+i \mathrm{sign}(\omega_{n})/2\tau}.
\end{equation}

\section{Superconducting instability at weak attraction}

\begin{figure}[t]  \centering
\includegraphics[width=8cm] {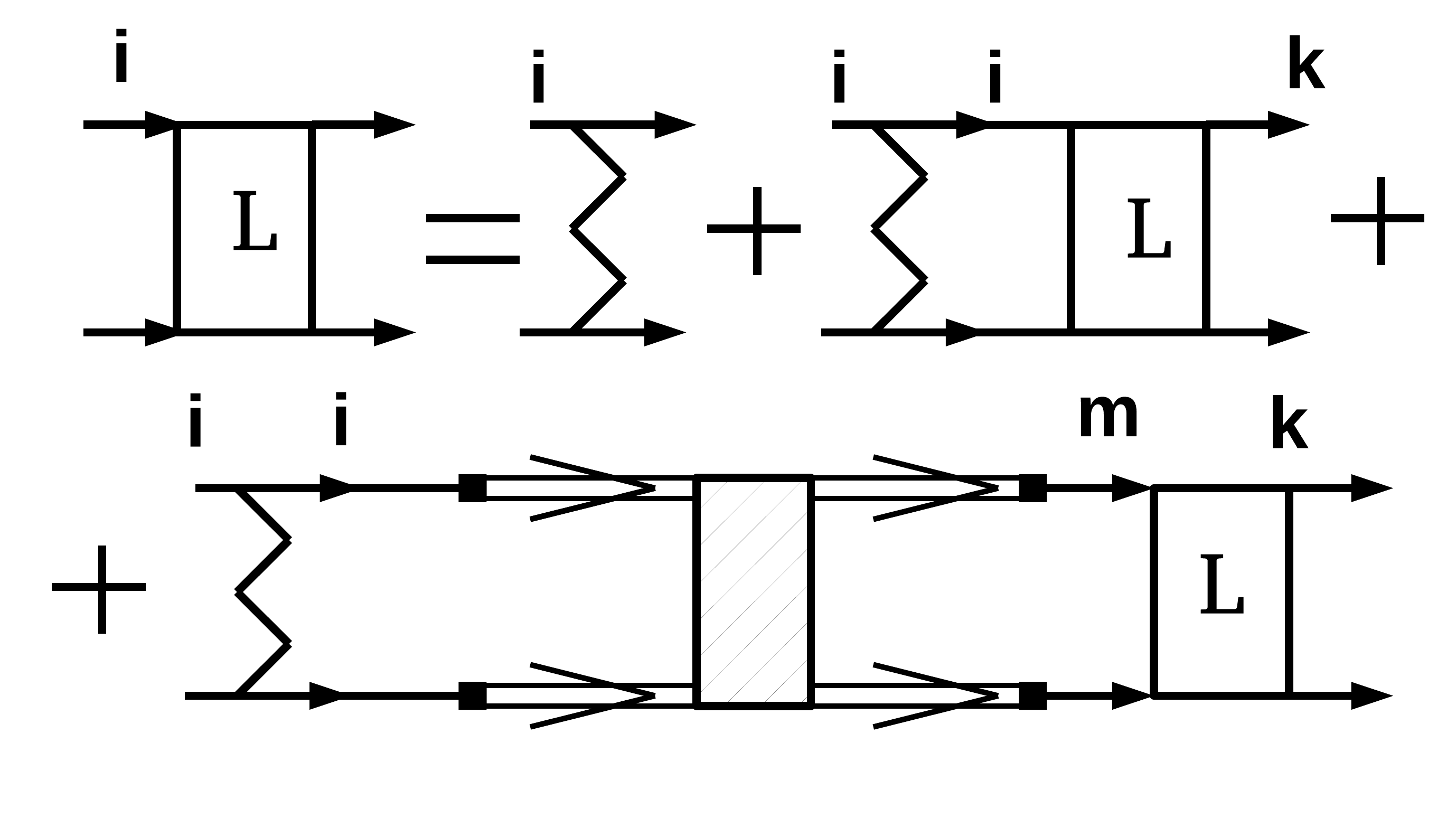}
\caption{Graphic equation for superconducting ladder $L(i,k)$, shown as rectangle with letter $\textsl{L} $. Single lines correspond to Green functions of resonance sites (\ref{g1-func}). Black square corresponds to hybridization to conducting band. Double line corresponds to Green function (\ref{cond-green}) of conducting electrons. Zigzag vertical lines correspond to on resonance site electron-electron interaction $W$. Propagation between sites $i$ and $m$ is described by the cooperon diffusion pole, shown in Fig.\ref{fig:2}. }\label{fig:1}
\end{figure}

\begin{figure}[t]  \centering
\includegraphics[width=8cm] {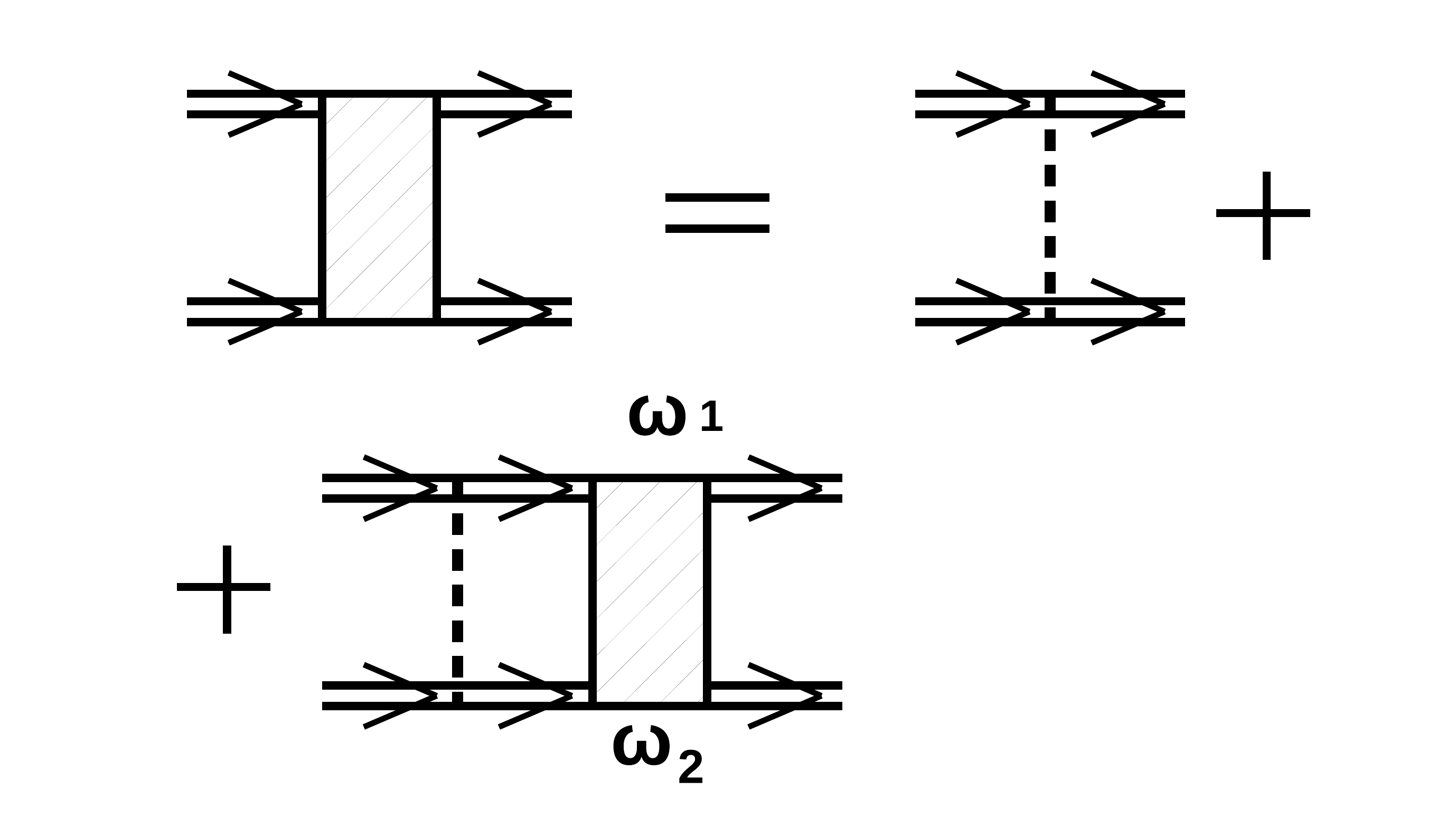}
\caption{Graphic equation for the cooperon-diffusion pole. Electron-impurity scattering is described by the vertical dotted line.}\label{fig:2}
\end{figure}

The instability of the system with respect to the transition to the superconducting state is described by the ladder $L(i,k)$, which is shown in Fig.\ref{fig:1}, \cite{bib:abricos}.
Note that we consider electron-electron interaction on the resonance sites only.
In the dilute system of resonance sites upon the propagation between them, conducting electrons experience multiple scattering events by impurities. 
These scattering processes can be described by the cooperon-diffusion ladder, which is schematically shown in Fig.\ref{fig:2} and is given by 
\begin{equation}\label{cooperon}
C(\textbf{q},\omega_{1},\omega_{2})=\frac{\Theta(-\omega_{1}\omega_{2})}{Dq^{2}+|\omega_{12}|},
\end{equation}
where $D$ is the diffusion constant of conducting electrons due to impurity scattering.

The ladder equation for $L(i,k)$ has discrete structure, where summation has to be performed over the resonance cites
\begin{equation}\label{ladder}
[1-W\Pi(0) ] L(i,k)=W\delta_{i,k}+W\sum_{i\neq m}\Pi(i-m)L(m,k).
\end{equation}

$W\Pi(0)$  is onsite contribution to ladder, determined by equation (\ref{BCS}).

Noting the long range nature of the cooperon-diffusion pole (\ref{cooperon}) in the definition of $\Pi(i-m)$, we replace the summation in (\ref{ladder}) over sites $k\neq i$ by the integration over position coordinate $\textbf{r}_{k}$ weighted with the density of resonance sites $n_{R}$. As a result we obtain
\begin{equation}
\sum_{i\neq m}\Pi(i-m)L(m,k)=n_{R}\int d\textbf{r}_{m}\Pi(\textbf{r}_{i}-\textbf{r}_{m})L(\textbf{r}_{m},\textbf{r}_{k}).
\end{equation}

In this approximation, the superconducting transition temperature is determined by equation 
\begin{equation}\label{weak-inst}
1-W\Pi(0)=Wn_{R}\int d\textbf{r}\Pi(\textbf{r}),
\end{equation}
where the right hand side is taken at zero momentum.

Dispersion equation (\ref{weak-inst}) can be written through the Green functions (\ref{g1-func}) in the form
\begin{align}\label{BCS}
&1-WT\sum_{\omega_{n}}g(\omega_{n})g(-\omega_{n})=\\\nonumber
&= \frac{2Wn_{R}\gamma^{2}}{\pi\nu_{0}}T\sum_{\omega_{n}} [g(\omega_{n})g(-\omega_{n})]^{2}C(\textbf{q}=0,\omega_{n},-\omega_{n}).
\end{align}
The right hand side in Eq. (\ref{BCS}) can be calculated as 
\[Wn_{R}\frac{\gamma^{2}\Lambda}{\pi^{2}\nu_{0}}\ln(E_{0}/T)\]
with logarithmic cutoff given by $E_{0}\sim \mathrm{min}(\gamma,|E_{R}|)$. Here we also introduce
\begin{equation}
\Lambda\equiv \frac{1}{(E_{R}^{2}+\gamma^{2})^{2}},
\end{equation}
which appeared due to the term $[g(\omega_{n})g(-\omega_{n})]^{2}$ taken in the limit $\omega_{n}\rightarrow 0$ and
will arise in all definitions of the transition temperature and critical magnetic fields.

The superconducting transition temperature can be found from Eq. (\ref{BCS}) in the weak coupling regime in the form
\begin{equation}\label{Weak Tc}
T_{c}=E_{0}\exp\bigg\{-\frac{\pi^{2}\nu_{0}(1-W/W_{c})}{\gamma^{2}\Lambda Wn_{R}}\bigg\},
\end{equation}
where we introduce
\begin{equation}\label{crit-attr}
W^{-1}_{c}=T\sum_{\omega_{n}}g(\omega_{n})g(-\omega_{n}).
\end{equation}
At $T \ll \gamma, |E_{R}|$ we find
\begin{equation}
W^{-1}_{c}=\frac{1}{2|E_{R}|}\left(1-\frac{2}{\pi}\arctan\frac{\gamma}{|E_{R}|}\right).
\end{equation}

Evidently, the weak coupling regime corresponds to $W/W_{c}<1$. In this case equation (\ref{weak-inst}) is satisfied due to the long-range nature of $\Pi(\textbf{r}_{i}-\textbf{r}_{m})$, which leads to the logarithmic contribution in (\ref{BCS}). At $W\ll W_{c}$ expression (\ref{Weak Tc}) resembles the one obtained in \cite{bib:Shelankov}.

Note that for positive $E_{R}>\gamma$, the strong coupling regime $W\sim W_{c}=2E_{R}$ corresponds the valence skipping condition \cite{bib:Varma}.

\section{Strong interaction $W/W_{c}\geq 1$}
It is convenient to consider the intermediate as well as strong coupling regimes within the framework of the Ginzburg-Landau (G-L) functional.
We separate the functional into the sum of local contributions of individual resonance sites and contributions, which are associated with the interaction between different sites.

\subsection{Local part of G-L functional}
The Green function, which is related to the Hubbard-Stratonovich transformation near the resonance site $i$, in saddle point approximation is given by
\begin{eqnarray}
g_{S,i}=-\frac{1}{\bar{\omega}_{n}^{2}+E^{2}_{R}+|\Delta_{i}|^{2}}
\begin{pmatrix}
 i\bar{\omega}_{n}+E_{R}&-\Delta_{i}\\
-\Delta^{*}_{i} & i\bar{\omega}_{n}-E_{R}
\end{pmatrix},~~~
\end{eqnarray}
where 
\begin{equation}
\bar{\omega}_{n}=\omega_{n}\left(1+\frac{\gamma}{|\omega_{n}|}\right).
\end{equation}
The self-consistent equation for the local $\Delta_{i}(\textbf{r})=\Delta_{i} |\psi(\textbf{r}-\textbf{r}_{i})|^{2}$ is given by
\begin{equation}
\Delta_{i}=WT\sum_{\omega_{n}}\frac{\Delta_{i}}{\bar{\omega}_{n}^{2}+E^{2}_{R}+|\Delta_{i}|^{2}}.
\end{equation}

This equation determines the dependence of $\Delta_{i}(T,W)$ on the temperature and attraction strength $W$. 
There is a critical value of attraction $W_{c}$, which is given by (\ref{crit-attr}) so that $\Delta_{i} (T,W)=0$ at $W<W_{c}$. 

The local part of G-L functional is given by a sum over the resonance sites
\begin{equation}\label{local}
F_{loc}=\sum_{i} (W^{-1}-W_{c}^{-1})\Delta^{2}_{i}+b\Delta^{4}_{i},
\end{equation}
where $W_{c}$ is determined by expression (\ref{crit-attr}), the coefficient $b$ is given by
\begin{equation}
b=T\sum_{\omega_{n}>0}\left[(\omega_{n}+\gamma)^{2}+E^{2}_{R}\right]^{-2},
\end{equation}
Note that both $W_{c}$ and $b$ are functions of $T$, $E_{R}$, and level broadening $\gamma$. To estimate, one can show that $b^{-1}\sim W^{3}_{c}\sim max (|E_{R}|^{3},\gamma^{3})$.

We would like to emphasize that in the strong coupling regime at $W>W_{c}$ the presence of nonzero value of $\Delta$ does not mean the existence of superconductivity. To have superconducting coherence we need the nonzero quantity
\begin{equation}
\langle\Delta_{i}\Delta_{k}\exp{i(\phi_{i}-\phi_{k})} \rangle\neq 0
\end{equation}
at large $|\textbf{r}_{i}-\textbf{r}_{k}|$ compared to the average distance between the centers.
This condition is determined by the nonlocal contribution to the functional G-L functional.

\subsection{Nonlocal part of G-L functional}

It is instructive to consider the nonlocal part of G-L functional as the result of Andreev reflections in a system with a given distribution of $\Delta_{i}$.

Using Nambu operators
\begin{equation}
\Psi(\textbf{r})=
\begin{pmatrix}
 \varphi_{\uparrow}(\textbf{r})\\
 \varphi^{+}_{\downarrow}(\textbf{r}) 
\end{pmatrix}, 
\Psi^{+}(\textbf{r})=
\begin{pmatrix}
 \varphi^{+}_{\uparrow}(\textbf{r})& \varphi_{\downarrow}(\textbf{r})
\end{pmatrix}
\end{equation}
the Hamiltonian of the system can be written as

\begin{eqnarray}
H_{\Delta} &=& H_\mathrm{cond} + \int d \textbf{r} \sum_{i} \bigg\{ E_{R}\Psi^{+}_{i}(\textbf{r})\tau_{z}\Psi_{i}(\textbf{r})-\\\nonumber
&-& \frac{1}{2}\Psi^{+}_{i}(\textbf{r}) 
[ \Delta_{i}(\textbf{r}) \tau_{x} + \Delta^{*}_{i}(\textbf{r})\tau_{y} ] \Psi_{i}(\textbf{r})\\\nonumber
&+& {V(\textbf{r})
[\Psi^{+}_{i}(\textbf{r})\tau_{z}\Psi(\textbf{r})+\Psi^{+}(\textbf{r})\tau_{z}\Psi_{i}(\textbf{r})} 
] \bigg\},
\end{eqnarray} 
where $H_\mathrm{cond}$ is the Hamiltonian of conducting electrons written in Nambu representation $\Psi(\textbf{r})$ and $\Psi^{+}(\textbf{r})$.

Let us consider a situation in which the order parameters is finite on sites $\textbf{r}_{i}$ and $\textbf{r}_{k}$ only.
In this case correction to the Hamiltonian of free electron gas is given by
\begin{equation}
\delta H=\delta H_{i}+\delta H_{k},
\end{equation}
where
\begin{equation}
\delta H_{i}=\int d\textbf{r}
\Psi^{+}(\textbf{r}-\textbf{r}_{i})\begin{pmatrix}
0& \Delta_{i}\\
\Delta^{*}_{i}& 0
\end{pmatrix}\Psi(\textbf{r}-\textbf{r}_{i}).
\end{equation}

\begin{figure}[t]  \centering
\includegraphics[width=8cm] {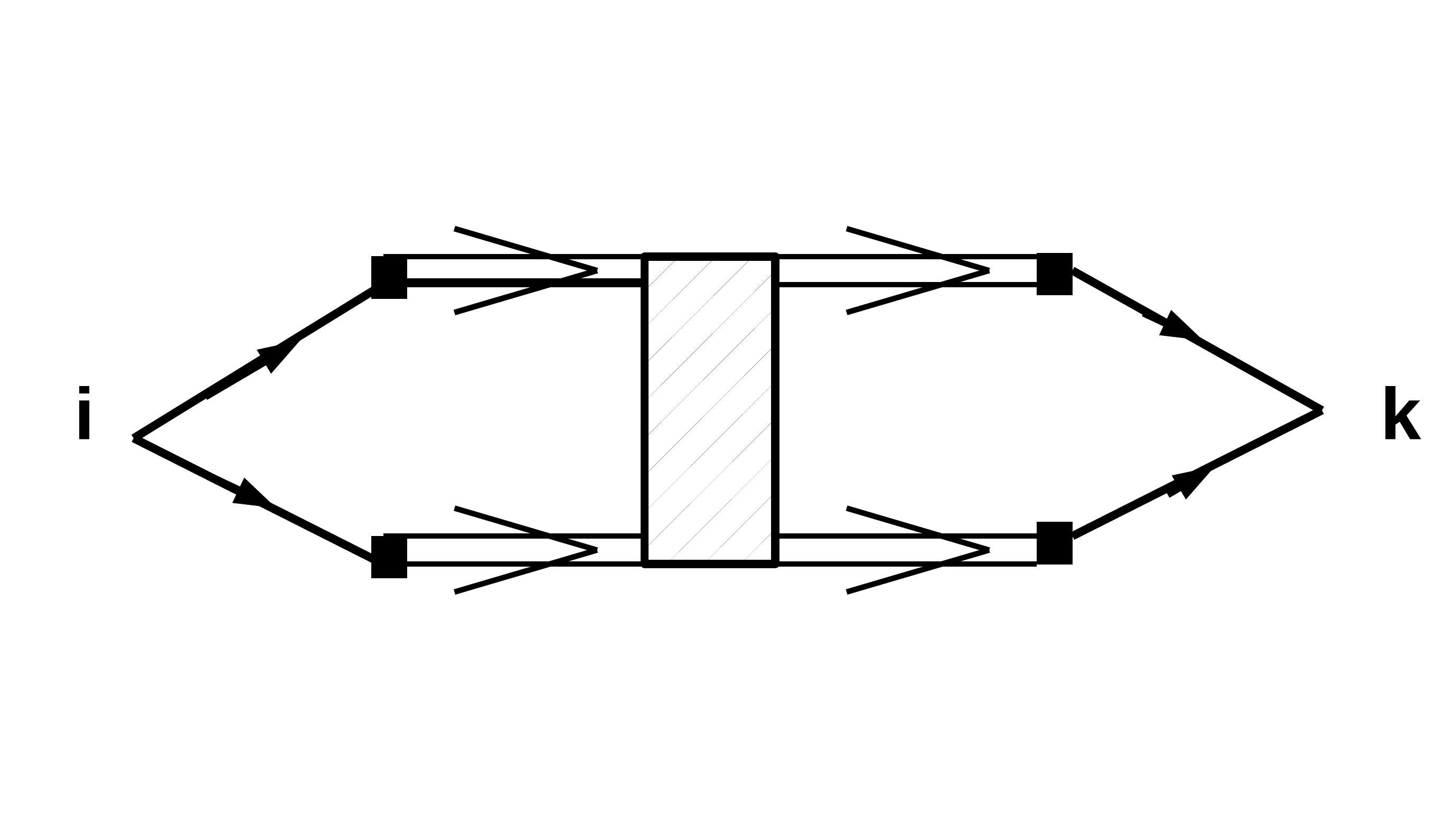}
\caption{Nonlocal correction to G-L functional. Multiple impurity scattering is described by the Cooper diffusion pole, shown on fig.\ref{fig:2}}\label{fig:3}
\end{figure}

The thermodynamic potential has a contribution, which is proportional to the product $\Delta_{i}\Delta_{k}$. This term corresponds to the diagram shown in Fig. \ref{fig:3} and is given by
\begin{equation}\label{correct}
F(i,k)=-T\int^{1/T}_{0}d\tau_{1}d\tau_{2} \langle T_{\tau}\delta H_{i}(\tau_{1})\delta H_{k}(\tau_{2})\rangle.
\end{equation}
Here the evolution of perturbation is determined as $\delta H_{i}(\tau)=\exp(\tau H_{\Delta=0})\delta H_{i}\exp(-\tau H_{\Delta=0})$.
Also note that expression in (\ref{correct}) depends of the superconducting phase difference $\phi_{ik}$.
One obtains

\begin{equation}\label{nonlocal}
F(i,k)=-B(i,k)|\Delta_{i}\Delta_{k}|\cos\phi_{ik},
\end{equation}
where 
\begin{eqnarray}\label{A-ik}\nonumber
B(i,k) &=& \frac{8\gamma^{2}}{\pi\nu_{0}}T\sum_{\omega_{n}>0}\int \frac{d\textbf{q}}{(2\pi)^{3}}\frac{\exp(i\textbf{qr}_{ik})}{2\omega_{n}+Dq^{2}}\\
&\times& \frac{1}{[(\omega_{n}+\gamma)^{2}+E^{2}_{R}]^{2}}.
\end{eqnarray}
The first factor in the denominator in (\ref{A-ik}) is due to the cooperon-diffusion ladder (\ref{cooperon}).

\subsection{Ginzburg-Landau functional at $W\sim W_{c}$}
Expressions (\ref{local}) and (\ref{nonlocal}) determine Ginzburg-Landau functional. One has
\begin{eqnarray}\label{G-L}\nonumber
F(\Delta_{i}e^{i\phi_{i}}) &=& \sum_{i}\bigg[\frac{1-W/W_{c}}{W}|\Delta_{i}|^{2}+b|\Delta_{i}|^{4}\bigg] \\
&-&\frac{1}{2}\sum_{i\neq k}B(i,k)|\Delta_{i}\Delta_{k}|\cos(\phi_{ik}).
\end{eqnarray}

\section{Temperature of transition to the coherent state}

Without magnetic field, the average value of the order parameter $\Theta\equiv \langle \Delta \rangle$ is a uniform real quantity.
Mean field equation for the complex order parameter $\Delta=\theta_{1}+i\theta_{2}$ in this case is given by
\begin{equation}\label{mean-f-defin}
\Theta\equiv \langle \theta_{1}+i\theta_{2}\rangle =
\frac{\int d\theta_{1}d\theta_{2}\Delta\exp(-F_{MF}(\Theta)/T)}{\int d\theta_{1}d\theta_{2}\exp(-F_{MF}(\Theta)/T)},
\end{equation}
where 
\begin{equation}
F_{MF}(\Theta)=\frac{1-W/W_{c}}{W}|\Delta|^{2}+b|\Delta|^{4}-\Theta\theta_{1}\sum_{k\neq 0}B(0,k).
\end{equation}

Expanding the right hand side of (\ref{mean-f-defin}) over $\Theta$, we obtain equation for the temperature of transition to the coherent state 
\begin{equation}\label{T-c}
1=\frac{ \langle |\Delta|^{2} \rangle}{2T}\sum_{k\neq 0}B(0,k),
\end{equation}
where 

\begin{equation}\label{delta}
\langle|\Delta|^{2} \rangle=\frac{\int^{\infty}_{0}  d\rho  \rho\exp \left[-\left(\frac{1-W/W_{c}}{W}\rho+b\rho^{2} \right)/T\right]}{\int^{\infty}_{0}  d\rho  \exp\left[-\left(\frac{1-W/W_{c}}{W}\rho+b\rho^{2} \right)/T\right]}.
\end{equation}

Performing a change $\sum_{k\neq 0}B(0,k)\rightarrow n_{R}\int d\textbf{r}B(\textbf{r})$, from (\ref{A-ik}) we obtain
\begin{equation}\label{A}
\sum_{k\neq 0}B(0,k)=\frac{2\gamma^{2}\Lambda n_{R}\ln(E_{0}/T)}{\pi^{2}\nu_{0}}.
\end{equation}

\begin{figure}[t]  \centering
\includegraphics[width=8cm] {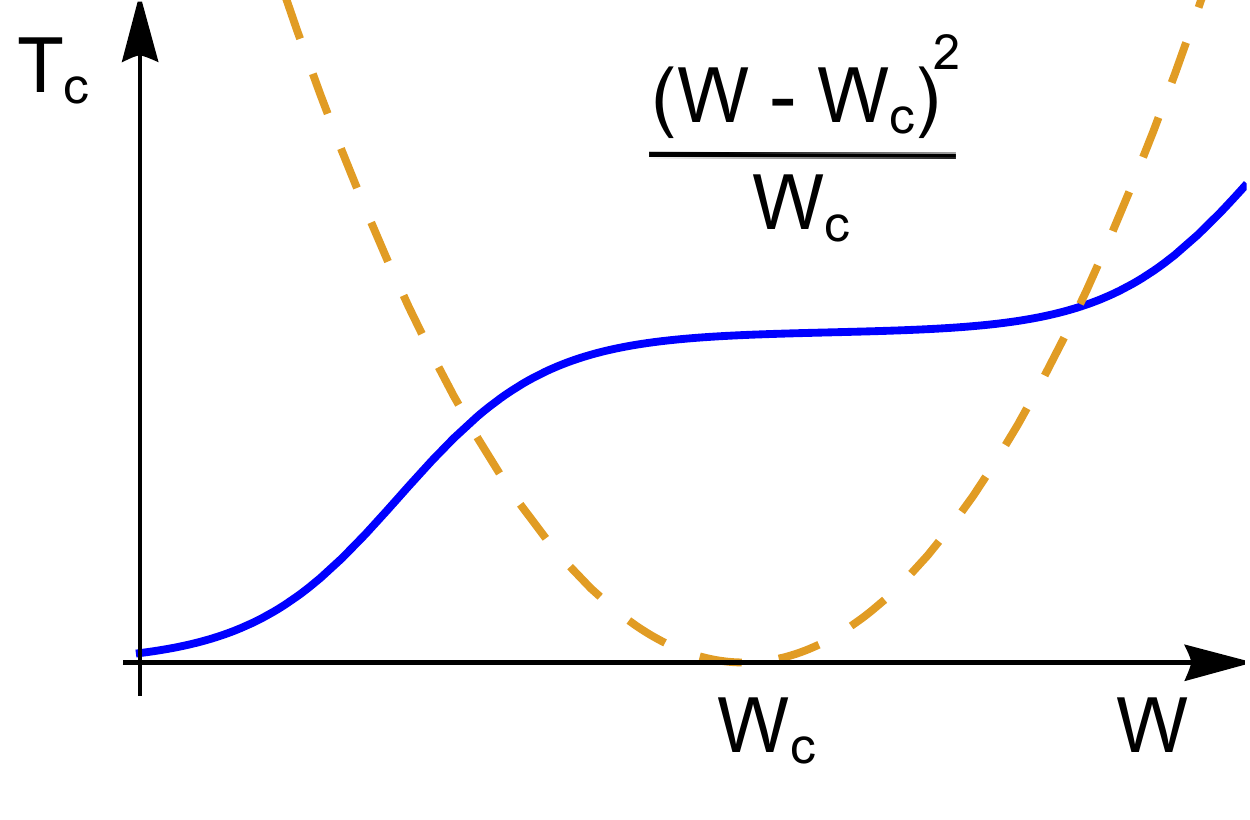}
\caption{Schematic dependence of the superconducting transition temperature on the attraction strength}\label{fig:4}
\end{figure}

By analyzing equation (\ref{T-c}), we obtain three regimes, which are schematically shown in Fig. (\ref{fig:4}).

Firstly, calculating correlation function in the weak coupling regime $W<W_{c}$ we can neglect the term $b\Delta^{4}$.
From (\ref{T-c}), (\ref{delta}), and (\ref{A}) we obtain the transition temperature in weak coupling limit (\ref{Weak Tc}).

Second, at $W\sim W_{c}$ and $T_{c}>(W-W_{c})^{2}/W_{c}$, we can neglect the term $\sim\rho$ in the exponents in (\ref{delta}).
In this case we obtain transition temperature from (\ref{T-c}) 
\begin{equation}\label{medium}
T_{c}=\bigg[ \frac{ n_{R}\gamma^{2}\Lambda}{\pi^{2}\nu_{0}\sqrt{\pi b}}\ln|E_{0}/T_{c}|\bigg]^{2}.
\end{equation}
This expression is valid for both $W<W_{c}$ and $W>W_{c}$.

Third, at stronger interaction the thermodynamic average of $\Delta$ has nonzero value. At $(W-W_{c})^{2}/W_{c}>T_{c}$, we can neglect fluctuation contribution in (\ref{delta}) so that 
the transition temperature is given by expression
\begin{equation}\label{large}
T_{c}=\frac{(W/W_{c}-1)}{Wb}
\frac{n_{R}\gamma^{2}\Lambda}{2\pi^{2}\nu_{0}} \ln|E_{0}/T_{c}|.
\end{equation}
We note that (\ref{large}) coincides with expression obtained in Ref. \cite{bib:Mal'shukov}.

At stronger interaction $W \gg W_{c}$, one gets $\langle |\Delta|^{2} \rangle = W^{2}$ in equation (\ref{T-c}). Therefore $T_{c}$ grows with $W$.

\section{Upper critical magnetic field}
Let us consider applied magnetic field with the vector potential $\textbf{A}(\textbf{r})$. In this case the average order parameter is nonuniform real quantity $\langle\Delta_{i} \rangle=\Theta_{i}$. 
Instead of (\ref{T-c}), the self-consistency equation is given by
\begin{equation}\label{H-c}
\Theta_{i}=\frac{\beta \langle |\Delta|^{2} \rangle}{2T}\sum_{k\neq i}B(i,k)\Theta_{k}.
\end{equation}
Replacing summation in Eq. (\ref{H-c}) with the integration and expanding the order parameter as 
\begin{eqnarray}
\Theta_{k}=\Theta(\textbf{r}_{i})+\frac{\textbf{r}^{2}_{ik}}{2} \left(\frac{d}{d\textbf{r}_{i}}-\frac{2e}{c}i\textbf{A}\right)^{2}\Theta(\textbf{r}_{i}) 
\end{eqnarray}
near $\textbf{r}_{i}$, 
on the right hand side of (\ref{H-c}) we obtain
\begin{eqnarray}\label{expansion}
\sum_{k\neq i}B(i,k)\Theta_{k} &=& n_{R}\int d\textbf{r} B(\textbf{r})\Theta(\textbf{r}_{i})+\\\nonumber
&+&\frac{n_{R}}{2}\int d\textbf{r} \textbf{r}^{2}B(\textbf{r}) 
\left(\frac{d}{d\textbf{r}_{i}}-\frac{2e}{c}i\textbf{A} \right)^{2}\Theta(\textbf{r}_{i}).
\end{eqnarray}

It is well known that in order to determine the upper critical magnetic field $H_{c2}$, it is necessary to choose an eigenstate $\Theta(\textbf{r})$ of (\ref{expansion}) with the the largest eigenvalue.
Finally, we find that the upper critical field is determined by equation
\begin{equation}\label{H-C-2}
n_{R}\int d\textbf{r} B(\textbf{r})-\frac{2T}{\langle |\Delta|^{2}\rangle}=\frac{n_{R}|e|H_{c2}}{c}\int d\textbf{r} \textbf{r}^{2}B(\textbf{r}). 
\end{equation}

According to (\ref{A-ik}) we have
\begin{equation}
\int d\textbf{r} \textbf{r}^{2}B(\textbf{r}) =\frac{D\Lambda\gamma^{2}}{2\pi\nu_{0}T_{c}}.
\end{equation}

In the weak coupling case $W< W_{c}$ and at temperature near $T_{c}$, we obtain
\begin{equation}
\frac{D|e|H_{c2}}{c}=\frac{4}{\pi}(T_{c}-T).
\end{equation}
At $W\sim W_{c}$, substituting  $\langle |\Delta|^{2} \rangle = (T/\pi b)^{1/2}$ into (\ref{H-C-2}), and for the system with the transition temperature (\ref{medium}) we obtain

\begin{equation}
\frac{D|e|H_{c2}}{c}=\frac{2}{\pi}(T_{c}-T)\ln(E_{0}/T_{c}).
\end{equation}

At strong attraction, when $W>W_{c}$ and $\langle |\Delta|^{2} \rangle = \langle |\Delta| \rangle^{2}= \left|\frac{1-W/W_{c}}{2bW} \right|$, the critical field can be found from 

\begin{equation}\label{strong field}
\frac{D|e|H_{c2}}{c}=\frac{4}{\pi}(T_{c}-T)\ln|E_{0}/T_{c}|.
\end{equation}

The critical field grows with the the attraction, approaching the value (\ref{strong field}) at $W>W_{c}$. 

We emphasize that the magnetic field does not affect the local value of $\Delta_{i}$. Rather, it reduces the effective values of the Josephson coupling between different sites.

\section{Conclusions}

In this paper we have studied the superconducting transition in dilute system of negative U-centers with strong electron-electron attraction.

We have shown that local densities of the Cooper pairs emerge at a temperature exceeding the superconducting transition temperature. 
The state exists provided attraction is stronger than some critical value $W_{c}$, which depends on the position and width of the resonance levels.
Global superconductivity is established due to the Andreev reflection between the resonance level at a lower temperature. This regime resembles the transition from the BCS state to the liquid of composite Bose particles.

Results do not change under changing sign of $E_{R}$. Therefore they are valid not only for valence skippers at small $|2E_{R}-W|<<E_{R}$ but also for system of sites with negative $E_{R}$ at $|2E_{R}+W|<<|E_{R}$

\section{Acknowledgment}
The author is thankful to A. L. Shelankov , V. Zyuzin and A. Zyuzin for illuminating discussions and Pirinem School of Theoretical Physics for hospitality.


\begin{thebibliography}{17}
\bibitem{bib:Leggett}
A. J. Leggett, J. Phys. (Paris) \textbf{41}, C7, 19 (1980).

\bibitem{bib:Nozieres}
P. Nozi\'eres, S. Schmitt-Rink,  J. Low Temp. Phys. \textbf{59}, 195 (1985).


\bibitem{bib:Melo}
C.A.R. S\'a de Melo, M. Randeria, J.R. Engelbrecht, 
Phys. Rev. Lett. \textbf{71}, 3202 (1993).


\bibitem{bib:Micnas}
R. Micnas, J. Ranninger, S. Robaszkiewicz
Rev. Mod. Phys. \textbf{62}, 113 (1990).


\bibitem{bib:Chernik}
I. A. Chernik, S. N. Lykov, 
Sov. Phys. Solid State \textbf{23}, 817 (1981).

\bibitem{bib:Kaidanov}
V. I. Kaidanov, Yu.I. Ravich, 
Sov. Phys. Usp. \textbf{28}, 31 (1985).


\bibitem{bib:Nemov}
S. A. Nemov and Y. I. Ravich, Sov. Phys. Usp. \textbf{41}, 735 (1998).

\bibitem{bib:Varma}
C. M. Varma, Phys. Rev. Lett. \textbf{61}, 2713 (1988).

\bibitem{bib:Mal'shukov}
A.G. Mal'shukov
Solid State Commun. \textbf{ 77}, 57 (1991).
and references therein. 

\bibitem{bib:Koba}
K. Kobayashi, Y. Ai, H.O. Jeschke, and J. Akimitsu1
Phys. Rev. B \textbf{97}, 104511 (2018).


\bibitem{bib:Haldol}
N. Haldolaarachchige, Q. Gibson, W. Xie, M. B. Nielsen, S. Kushwaha, and R. J. Cava,
Phys. Rev. B \textbf{93}, 024520 (2016).


 \bibitem{bib:Girado-Gallo}
P. Girado-Gallo et al., Phys. Rev. Lett. \textbf{121}, 207001 (2018).

\bibitem{bib:kriener}
M. Kriener, M. Sakano, M. Kamitani, M.S. Bahramy, R. Yukawa, K. Horiba, H. Kumigashira, 
K. Ishizaka, Y. Tokura, and Y. Taguchi, Phys. Rev. Lett. \textbf{124}, 047002 (2020).


\bibitem{bib:Dzero}
M. Dzero, J. Schmalian, 
Phys. Rev. Lett. \textbf{94}, 157003 (2005).


\bibitem{bib:Matsushita}
Y. Matsushita, H. Bluhm, T. H. Geballe, and I. R. Fisher,
Phys. Rev. Lett. \textbf{94}, 157002 (2005).

\bibitem{bib:abricos}
A. A. Abrikosov, L. P. Gor'kov, and I. E. Dzyaloshinskii,
"Methods of Quantum Field Theory in Statistical Physics"
(Dover, New York, 1963).
 
\bibitem{bib:Shelankov}
A. L. Shelankov, 
Solid State Commun. \textbf{62}, 327 (1987).

\end{thebibliography}
\end{document}